\documentclass[useAMS,usenatbib]{mn2e}

\usepackage {graphicx}
\usepackage {times}
\usepackage {amssymb}
\usepackage {multirow}
\usepackage {color}

\title[MIR LFs  of the NEP-Wide Field]
       {Mid-Infrared Luminosity Function of Local  Star-Forming Galaxies 
            in the NEP-Wide Survey Field of \textit{AKARI} }

\author[S. J. Kim et al.]{Seong Jin Kim,$^{1,2}$\thanks{E-mail: seongini@kasi.re.kr}
           Hyung Mok Lee,$^{1}$   
           Woong-Seob Jeong,$^{2,3}$
          Tomotsugu Goto,$^{4}$   \newauthor
          Hideo Matsuhara,$^{5}$ 
          Myungshin Im,$^{1,6}$ 
         Hyunjin Shim,$^{7}$  
         Min Gyu Kim$^{1}$ and \newauthor
         Myung Gyoon Lee$ ^{1}$  \\
        $^{1}$Astronomy Program, FPRD, Department of Physics and Astronomy, SNU,  
         Kwanak-Gu, Seoul 151-742,  Korea;
        seongini@astro.snu.ac.kr \\
       $^{2}$ Korea Astronomical Space Science Institute, 61-1, Whaam-dong, Yuseong-gu, 
        Deajeon 305-348, Republic of Korea; seongini@kasi.re.kr\\
       $^{3}$  Korea University of Science and Technology, 217 Gajeong-ro, Yuseong-gu, 
               Deajeon 305-350, Republic of Korea \\       
       $^{4}$Institute of Astronomy, National Tsing Hua University, No. 101, Section 2, 
               Kuang-Fu Road, Hsinchu, Taiwan 30013, R.O.C\\
       $^{5}$Institute of Space and Astronautcial Science, Japan Aerospace
        Exploration Agency, Yoshinodai 3-1-1, Sagamihara, Kanagawa 229 8510\\
       $^{6}$Center for the Exploration of the Origin of the Universe (CEOU),
        Seoul National University, Seoul 151-742, Republic of Korea \\
       $^{7}$Department of Earth Science Education, Kyungpook National University, 
         Deagu 702-701, Republic of Korea }
         
\date{Accepted 2015 Aug xx; Received 2015 May xx; in original form 2015 May xx}
\pagerange{\pageref{firstpage}--\pageref{lastpage}} \pubyear{2015}

\begin{document}

\maketitle

\label{firstpage}

\begin{abstract}

We present mid-infrared (MIR) luminosity functions (LFs) of local ($z < 0.3$) star-forming
(SF) galaxies in the \textit{AKARI}'s NEP-Wide Survey field. In order to derive more accurate
luminosity function, we used spectroscopic sample only.  Based on  the NEP-Wide point 
source catalogue containing a large number of infrared (IR) sources distributed over the wide
(5.4 deg$^2$) field,  we incorporated the spectroscopic redshift ($z$) data for $\sim$ 1790
selected targets obtained by optical follow-up surveys with MMT/Hectospec  and 
WIYN/Hydra.  The \textit{AKARI}'s continuous 2 -- 24$\mu$m wavelength coverage as
well as photometric data from optical  $u^{*}$ band to  near-infrared $H-$band
with  the spectroscopic redshifts for our sample galaxies enable us to derive accurate spectral 
energy distributions (SEDs) in the mid-infrared. 
We carried out SED-fit analysis and employed $1/V_{max}$ method to derive the MIR 
(e.g., 8 $\mu$m, 12 $\mu$m, and 15 $\mu$m rest-frame) luminosity functions. 
We fit our 8  $\mu$m LFs to the double power-law with the power index of $\alpha = 1.53$
and $\beta = 2.85$  at the break luminosity $4.95 \times 10^{9} L_{\odot} $.  We 
made extensive comparisons with various MIR LFs  from several literatures. Our results for
local galaxies from the NEP region are generally consistent with other works for different fields
over wide luminosity ranges. The comparisons with the results from the NEP-Deep data as well
as other LFs imply the luminosity evolution from higher redshifts towards the present epoch.

\end{abstract}


\section{INTRODUCTION}

The Luminosity Function (LF), number of galaxies per unit volume per unit luminosity bin, is an important 
indicator of the distribution of galaxies over cosmological time.  In particular, infrared (IR) LFs provide useful
 information regarding the amount of energy released by galaxy activities. We can obtain, however, 
different information depending on the wavelength at which  the LF is measured.  For example, NIR 
luminosity function should be regarded as a proxy of stellar mass function of galaxies,  while various 
emission features in the MIR provide clues to  detailed dust properties related to the star-forming activities. 
 Dust emission peaks in the far-IR (FIR) wavelengths and FIR luminosities give insight to total infrared 
 luminosity ($L_{IR}$) which  is directly connected to overall bulk of energy emitted by star-formation 
 activities. 

The construction of the infrared luminosity functions has begun based on the data obtained by \textit 
{Infrared Astronomical Satellite} (IRAS) in the mid- and far-infrared,  although the sample  was restricted 
to bright galaxies because of relatively low sensitivity of the IRAS (Rowan-Robinson,  Helou and Walker 
1987; Saunders et al. 1990; Rowan-Robinson et al. 1997).   Subsequently, \textit {Infrared Space 
observatory} (ISO) data was used to determine the mid-infrared (12 $\mu$m and 15$\mu$m) LFs  
with follow-up imaging and spectroscopic observations (Clements et al.  2001; Xu et al. 2000; Pozzi 
et al. 2004). Far-infrared  LF was also estimated (e.g., 90 $\mu$m LF by Serjeant et al. 2004) using 
the European Large-Area ISO Survey (ELAIS) final (optical-IR merged) catalogue (Rowan-Robinson et al.
2004).  Based on the IRAS (Hacking et al. 1987; Franceschini et al. 1988; Fang et al. 1998 etc.) and 
ISO data (Elbaz et al. 1999; Puget et al. 1999), it has been revealed that dust-enshrouded star-forming
galaxies  undergo evolution in luminosity and density up to redshift around $z \sim 1$, and that their 
evolutionary rate may exceed those measured at any other wavelengths. The contribution of infrared 
luminous galaxies (such as LIRGs and ULIRGs defined as $L_{IR} > L_{\odot}^{11}$ and $L_{IR} > 
L_{\odot}^{12}$, respectively ) to the star formation (SF) history has been recognized and well 
established recently, implying the star forming activity was much more intense at higher redshifts ($z$) 
than nearby universe.  

More accurate infrared luminosity functions (IRLFs) have been derived for both local and distant galaxies
observed from various fields  based on the \textit{AKARI} and  \textit{Spitzer} data by many authors
(P{\'{e}}rez-Gonz{\'a}lez et al. 2005; Babbedge et al. 2006; Caputi et al. 2007; Goto et al. 2010;
Rodighiero et al. 2010; Goto et al. 2011; Wu et al. 2011).   
They have found a positive evolution in luminosity and density and increasing importance of the LIRGs and 
ULIRGs populations toward higher redshifts, indicating that the universe was  much more active  in the 
past.  Recent interests have moved to even higher redshift  beyond the peak ($z \sim 2$) of the cosmic
 SFR density (e.g., P{\'{e}}rez-Gonz{\'a}lez et al. 2005; Rodighiero et al. 2010; Magnelli et al. 
2011)  to understand  the early history  of the star formation. However,  we still have many questions
about the local universe and uncertainties in the parameters that can relate local galaxy formation and 
evolution  with the global history of the universe.  Especially, accurate measurements of the LFs in the 
local  universe is important in order to determine the course of evolution over cosmic time. 

\textit{AKARI}, an IR mission by JAXA/ISAS in Japan with European and Korean collaborations, has
carried out all sky surveys at mid- and far-infrared (Murakami et al. 2007). \textit{AKARI} also 
carried out a number of wide area surveys with better sensitivities than all-sky survey, including the
North Ecliptic Pole (NEP) surveys which used continuous nine spectral bands ranging from 2 to 24
$\mu$m. The NEP survey was composed of two programs: Wide and Deep surveys. The NEP-Deep
survey covered about 0.4 deg$^2$ field with higher sensitivity (Matsuhara et al. 2006; Wada et al. 
2008; Takagi et al. 2012), while the NEP-Wide survey covered much wider area of about 5.4 
deg$^2$ with lower sensitivity. The NEP-Wide survey is about 0.5--0.6 mag shallower than NEP-Deep
survey in the NIR and MIR-S bands. The scientific purpose and overall plan of the NEP survey projects
were presented by Matsuhara et al. (2006), and detailed description on the NEP-Wide and NEP-Deep
 surveys including their IR source catalogues have been published by Kim et al. (2012) and Takagi et 
 al (2012), respectively.
 
 Based on the data obtained by NEP-Deep survey, Goto et al. (2010, hereafter referred to as G10) have 
 derived the MIR (8 and 12 $\mu$m) luminosity functions for galaxies at redshift  from  0.3 to  2. 
They used $\sim$ 4,000 IR sources and estimated photometric redshifts to determine the distances to
galaxies based on SED model fitting. Recently, they updated their LFs using newly obtained CFHT data 
(Goto et al. 2015).   In this work, we use the NEP-Wide Survey data which has more galaxy sample in the 
local universe,  where G10 did not cover.  The primary purpose of this work is to present more 
accurate  luminosity functions for the redshift $z < 0.3$  using only spectroscopic sample in the NEP-Wide 
field.  Therefore, our study complements to G10  in the following aspects.  
First, we use the NEP-Wide survey data covering the wider angular coverage (5.4 deg$^2$) that 
allows us to minimize the cosmic variance. Second, since the photometric depth of the NEP-Wide survey
is shallower,  we concentrate on the derivation of the luminosity function for the local universe within 
$z \sim 0.3$,  whereas G10 measured  the 8$\mu$m and 12$\mu$m LFs  for the
redshift range  $z > 0.3$.  Finally, we take advantage of the
spectroscopic  redshifts (spec-z) for  galaxies in the NEP-Wide field obtained by the follow-up
spectroscopic surveys (Shim et al. 2013) with  MMT/Hectospec and WIYN/Hydra. 
Thus, our estimation of the distances  to the galaxies would be accurate.

This paper is organized as follows. In section 2, we present IR photometric  data from \textit{AKARI} 
NEP-Wide survey and spectroscopic redshifts (spec-$z$) data from optical survey with MMT/Hectospec
and WIYN/Hydra used in this study. In section 3, we describe our methodology to build the LFs.  We 
present our results in section 4,  and conclusions and summary in section 5. Throughout this paper, 
we assume a $\Lambda$ cold dark matter ($\Lambda$CDM) cosmology with $H_{0} =$ 
70 km s$^{-1}$ Mpc$^{-1}$, $\Omega_{m} =$ 0.3, $\Omega_{\lambda} =$ 0.7.

\section{DATA and ANALYSIS }

\subsection{Multi-wavelength NEP-Wide data}

The NEP-Wide Survey (Matsuhara et al. 2006; Lee et al. 2009; Kim et al. 2012) is one of the large 
area survey programs of \textit{AKARI} space telescope (Murakami et al. 2007).  The imaging survey 
toward the north ecliptic pole (NEP) was carried out on the circular area of about 5.4 deg$^2$ with nine
photometric  bands of the Infrared Camera (IRC, Onaka et al.  2007), covering the spectral range of 2
-- 24 $\mu$m nearly continuously. The photometric bands are designated as $N2$, $N3$, and $N4$
for the NIR, $S7$, $S9W$, and $S11$  for the MIR-S (shorter part of the MIR) bands, and $L15$, 
$L18W$, and $L24$ for the MIR-L (longer part of the MIR) bands: the numbers indicate approximate
effective wavelength in units of $\mu$m.

 The detailed description of the data reduction methodologies, catalogue of point sources and their 
characteristics are presented by Kim et al. (2012).  Here we only make brief summary of the NEP-Wide
data sets.  All magnitudes of IR sources in this work are reported in units of AB, hereafter.  The entire 
region surveyed by the NEP-Wide program centered at the NEP ($\alpha =18^{h}00^{m}00^{s}$, 
$\delta = 66^{\circ}33^{'}38^{''}$) can be found in Figure 1 (left panel).  Background gray tiles in this 
figure represent 446 pointed observations carried out using IRC instruments  to cover the whole NEP field.  
The 5-$\sigma$ detection limits of NEP-Wide survey are approximately 21 magnitude (mag) in the  
NIR bands, 19.5 -- 19 mag in the MIR-S bands, and 18.8 -- 18.5 mag in the MIR-L bands in AB 
mag, respectively. The NEP-Wide point source catalogue contains more than $\sim$ 100,000 sources
in the NIR and up to $\sim$ 18,000 sources in the MIR bands.  Compared to the  detection limits, 
 50\% completeness levels are 0.5 -- 0.6 mag brighter in the MIR and $\sim$ 1.2 mag brighter in 
 the near-IR bands.  The depth of the NIR bands are affected by source confusion while those of the 
 MIR bands are less affected by confusion because of the lower source density. The number of 
sources in the MIR bands typically ranges from a few thousands to about eighteen thousands.

\begin{figure*}
\begin{center}
\includegraphics[scale=0.55]{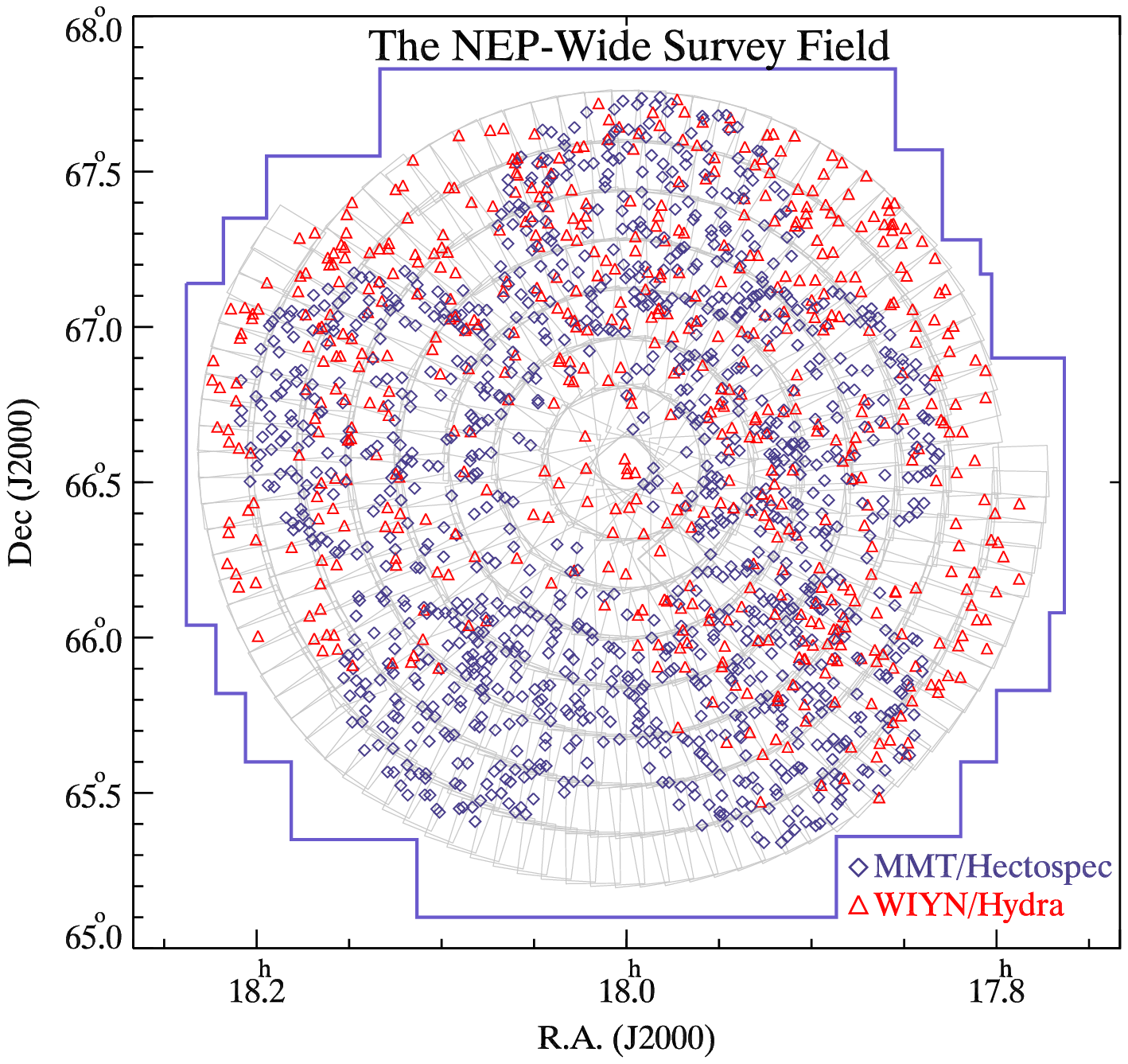}
\includegraphics[scale=0.55]{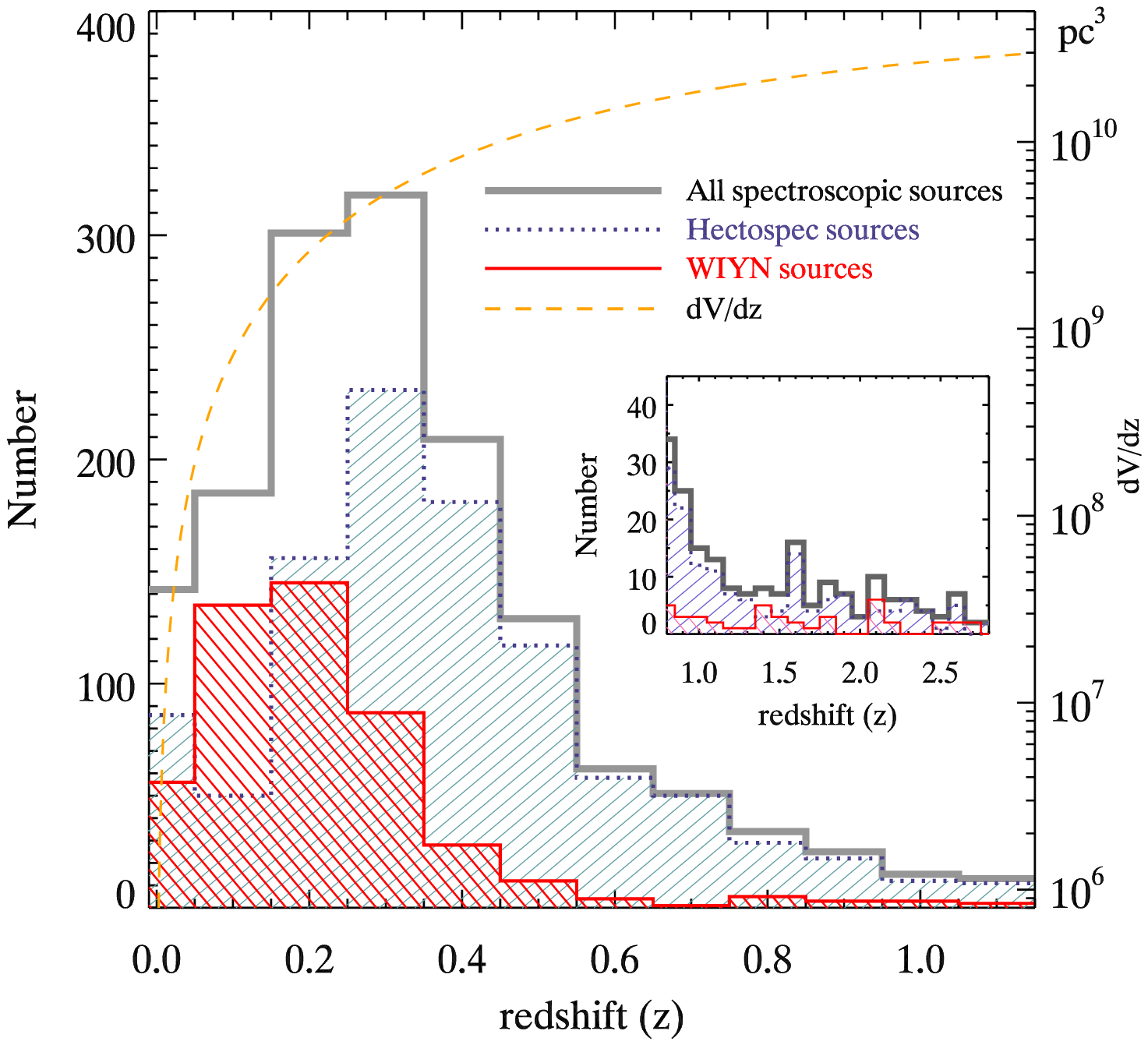}
\end{center}
\caption{Left panel shows the spatial distribution of the sources observed by follow-up spectroscopic 
surveys over the NEP-Wide field.  The centre of this map is the North Ecliptic Pole ($\alpha =
18^{h}00^{m}00^{s}$, $\delta= +66^{\circ} 33^{'}38^{''} $). Background gray boxes represent
446 pointed observation of IRC towards this field.  Right panel shows the number of sources as a 
function of redshift.  The blue colour indicates the spectroscopic targets observed using Hectospec, and 
the red colour shows the sources observed using WIYN. Also shown with yellow curve is $dV_{z}/dz$ 
in arbitrary units.}
\label{fig 1}
\end{figure*}

 One of the advantages of using the NEP-Wide data is that we have many ancillary data sets such as 
 high-quality optical data obtained with the MegaCam of 3.5m CFHT (Hwang et al. 2007) and the 4k 
 $\times$ 4k  SNUCAM (Im et al. 2010) of the  1.5m telescope at Maidanak observatory in Uzbekistan 
 (Jeon et al. 2010).  In addition, near-IR $J-$, $H-$ band data have been obtained (Jeon et al. 2014) 
 using the FLoridA Multi-object Imaging Near-ir Grism Observational Spectrometer (FLAMINGOS; Elston
 et al. 2006) mounted at the Kitt Peak National Observatory (KPNO) 2.1m telescope, and various radio
data over a limited region (Kollgaard et al. 1994, Lacy et al. 1995, Sedgwick et al. 2009, White et al. 
2010).   Recently, the \textit{Herschel Space Observatory} (Pilbratt et al., 2010) covered the entire 
\textit{AKARI}  NEP-Wide field with the Spectral and Photometric Imaging Receiver (SPIRE; Griffin et al. 
2010).

For the IR sources detected by \textit{AKARI} NEP-Wide survey, all the photometry results of IRC bands 
as well as ancillary data were produced in the form of (optical to MIR) band-merged catalogue covering  
$u^{*}$ band to  24 $\mu$m band (Kim et al. 2012)\footnote{See also  
http://www.ir.isas.jaxa.jp/AKARI/Archive/Catalogues/NEPW\_V1/}.  To construct a reliable source 
catalogue, all of the detected sources at one \textit{AKARI}/IRC band were cross-matched with those 
in the other IRC bands and all the ancillary data sets including optical (CFHT and Maidanak) and 
FLAMINGOS near-IR bands ($J$ and $H$). The catalog contains about {114,800} sources detected
 at least in one of the IRC filter bands. No attempt to distinguish the sources into different types such as 
 stars, galaxies has been made yet systematically for the entire sources in this NEP-Wide Catalogue. 
 However, we inspected sub-sample of NEP sources  based on various colour-colour diagrams with the 
 stellarity parameters of optical counterparts, and  assessed that about 70\% of the MIR sources 
 (with $S11$  magnitude brighter than 18.5 mag) are star-forming galaxies mostly at redshift range  
 z $<$ 1 (see Lee et al. 2007; Lee et al. 2009; Kim et al. 2012). The active galactic nuclei (AGNs) 
 and early-type galaxies comprise approximately 10\% or more   among  the sample we inspected.  
 The rest of them are most likely galactic stars.  We expect that the composition of entire MIR sources 
 in the NEP-Wide catalogue should be similar. 

\subsection{Spectroscopic Data and Galaxy sample}

Follow-up spectroscopic surveys over the entire \textit{AKARI} NEP-Wide field were carried out with 
MMT/Hectospec and WIYN/Hydra (Shim et al. 2013).   The targets for spectroscopic follow-up 
observation were selected primarily based on MIR fluxes at 11 $\mu$m ($<$ 18.5 mag, or $f_{S11} 
>$ 150 $\mu$Jy)  and at 15 $\mu$m ($<$ 17.9 mag, or $f_{L15} >$ 250 $\mu$Jy). An additional
$R$-band magnitude cut was imposed to select optically bright objects ($16 < R < 21$ for Hydra, $16 
< R < 22.5$ for Hectospec observations) that can yield the spectra with reasonable signal-to-noise 
(S/N) ratio. Most of these flux-limited targets are considered to be various types of IR luminous 
star-forming galaxies.  In addition to these primary targets,  a smaller number of secondary targets were
selected using their optical colours ($g - r$, $B - R$, etc.) and IRC band colours ($N2 - N4$, $S7 - 
S11$). These secondary criteria include high-redshift galaxy candidates based on dropout techniques 
($u$-dropout for $z \sim$ 3 and $g-$ or $B-$dropouts for z $\sim$ 4)\footnote{10 sources are 
not selected from the NEP-Wide catalogue}, active galactic nuclei (AGN) candidates selected by the NIR
($N2 - N4 > 0$) and MIR ($S7 - S11 > 0$) colours reflecting power-law SED of an AGN (e.g., Lee 
et al. 2007), radio sources (White et al. 2010), polycyclic aromatic hydrocarbon (PAH)-luminous 
galaxies (Ohyama et al.,2009; Takagi et al., 2010), and so on. 
The supercluster member candidates  in the NEP field at z $=$ 0.087 (Ko et al. 2012) were 
included in the  targets for this spectroscopic observation. Majority of the secondary 
targets are not used  in the analysis for the LF derivation  because they are cluster members  or at 
$z > 0.3$.

\begin{figure*}
\begin{center}
\includegraphics{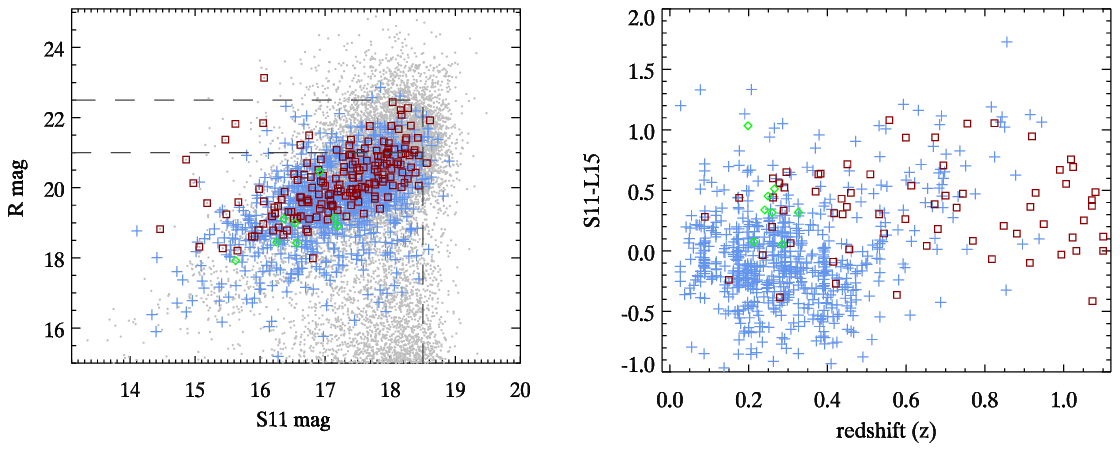}
\includegraphics{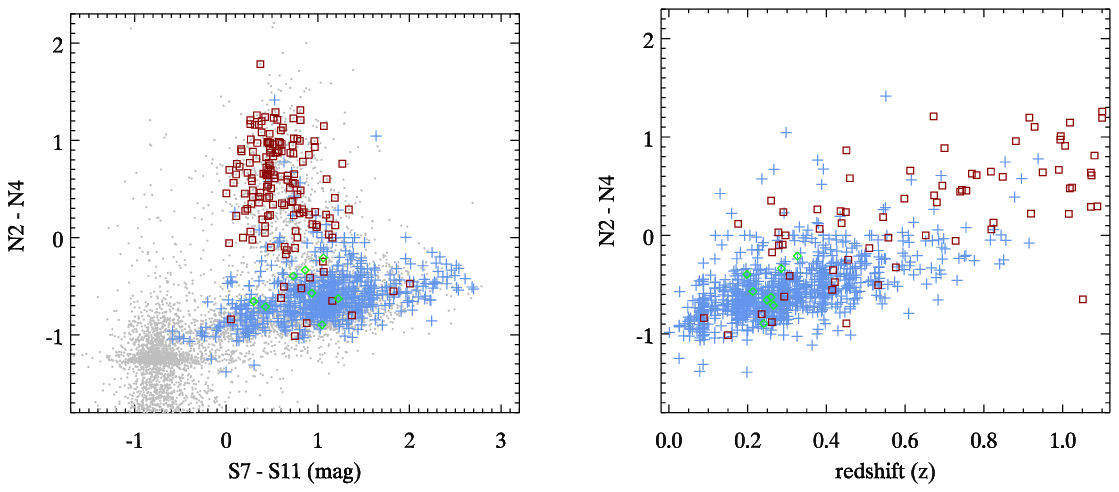}
\end{center}
\caption{The upper left panel shows the distribution of spectroscopic sample in  optical $R$  band  versus
 MIR  $S11$ band magnitudes plane, together with the primary selection criteria from $S11$ band 
and additional  $R$-band magnitude cut for spectroscopic observations.  Upper right panel shows the
distribution of $S11-L15$ colour as a function of redshift.  Lower left panel is an near- and mid-infrared
colour-colour  diagram showing photometric properties of the spectroscopic sample, and lower right panel 
shows the NIR colour distribution as a function of redshift.  Backgroundnd gray dots represent parent 
photometric sample, and cross indicates sample classified as `galaxy'.   Square and diamond indicate  those 
classified as type-1 and type-2 AGNs, respectively.   }
\label{fig 2}
\end{figure*}

Shim et al. (2013) presented spectra of 1796 sources (primary targets: 1155, secondary targets: 
641), among which spectroscopic redshifts are  measured for 1645 objects.  The success rate for 
redshifts identification are higher than 80\% at R $<$ 21 mag, though the quality of the spectra varies
significantly depending on the targets in different fields. Quality flag ranging from 1 to 4 was assigned to
each source: 4 for a secure redshift, 3 for an acceptable and almost good redshift, 2 for a questionable,
and 1 for unusable  redshift.  We did not use the sources having the flag 1 or 2 ($\sim$ 190).  The 
spectroscopic redshifts of $\sim$ 90\% of the sources are distributed over the redshift range of z $<$
1.0 (120 sources out of these are galactic stars), and the remaining 10\% lie at z $>$ 1.0.  Due to 
the target selection seeking for high-redshift AGNs, most sources at z $>$ 1.0 are  Type-1 AGNs. 
Details of the line flux measurements and diagnostic line ratios can be found in  Shim et al. (2013). 
The spatial distribution of the NEP-Wide sources  observed by this spectroscopic survey is presented  in
 the left panel of Figure 1 along with the NEP-Wide survey area.  The blue diamonds indicate the
spectroscopic targets observed with MMT/Hectospec, and the red triangles show the targets observed
with WIYN/Hydra. We also show the redshift distribution of this spectroscopic sample in the right
panel of Figure 1.  All the spectroscopic  sources are shown in black colour.  
The blue line indicates the distribution of sample observed with MMT/Hectospec, and the red one
represents that observed with WIYN/Hydra. Also shown with a yellow curve is the shape of $dV_{z}/dz$,
where $V_{z}$ is the comoving volume  within redshift $z$. It is clear that our spectroscopic sample 
seriously underrepresents the galaxies beyond  $z > 0.35$,  but is consistent  with  constant spatial 
density below $z < 0.35$.  We used the sample at $z < 0.3$, where the spectroscopic incompleteness 
is less severe.  The redshift distribution of the sources with higher redshifts ($> 1$) is shown in an inner 
small box.

\begin{figure*}
\includegraphics[scale=0.8]{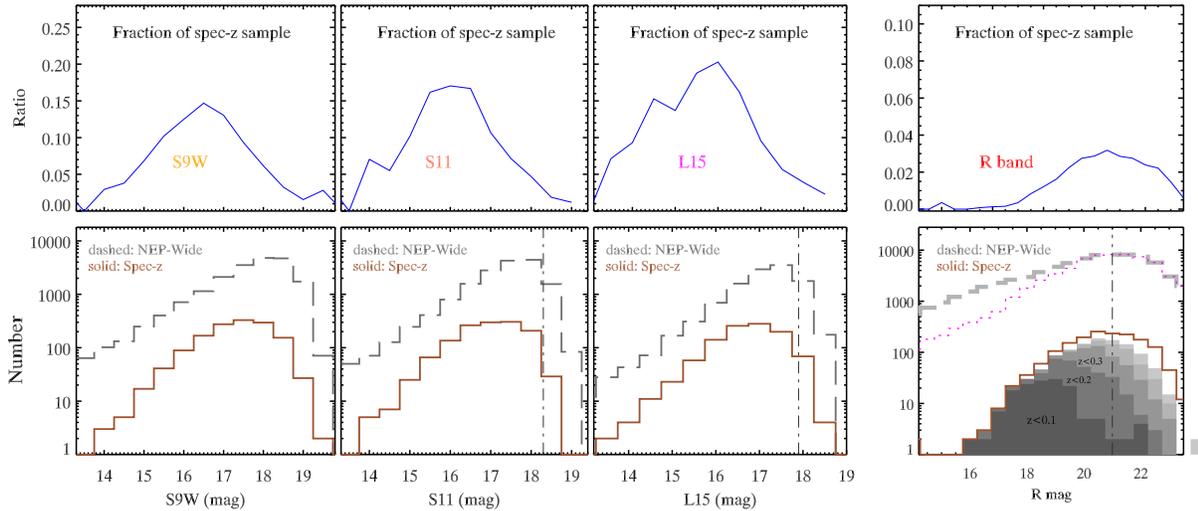}
\caption{Upper panels show the fraction of spectroscopic sample relative to the photometric   
sources as  a  function of observed magnitudes at the MIR bands of \textit{AKARI} ($S9W$, $S11$, 
and $L15$) and optical R band (the rightmost panels).  In the lower panels, dashed line indicates
the number  of parent  photometric sources and solid line indicates the spectroscopic sample.
Dotted line in the rightmost panel indicates the sources except for the stars.  Vertical dot-dashed lines indicate magnitude cut  used for the target selection. 
\label{fig 3}}
\end{figure*}

In Figure 2, we present the IR properties of the parent photometric sources and spectroscopic sample used
in this work.  Using analysis of emission line fluxes, Shim et al. (2013) identified the types of spectroscopic 
sources. They presented Baldwin-Phillips-Terlevich (BPT) diagram for the sources using emission lines 
ratios such as [O III]/H$\beta$  and [NII]/H$\alpha$, and classified them into star-forming galaxies,
starburst-AGN composites, and AGNs.  The sources classified as `galaxy' are indicated by crosses 
(blue colour) while those classified  as `AGN' types are indicated by small boxes (squares for type-1  and diamonds for type-2), whose types are identified based on the 
analysis of emission lines (Shim et al. 2013).    In the upper left panel of Figure 2,  we give the 
distribution of the optical ($R$) versus MIR ($S11$) band magnitudes of the NEP-Wide sources, 
together with the primary criteria from $S11$ band and additional $R$-band magnitude cut  for 
the spectroscopy. The background gray dots indicate parent photometric sources from the NEP-Wide 
catalogue.  Upper right panel shows the distribution of MIR-band colour  $S11 - L15$ as a function of 
redshift, showing most of the galaxy sample gathered between -0.5 and 0.5  in this colour range with 
the redshift $z < 0.5$,  compared to the wider spread of the AGN types.  
In the lower left panel, the colour-colour diagram of the \textit{AKARI} NIR and MIR bands  shows that 
the sources classified as AGNs are clearly distinguished by the relatively red NIR colour ($N2-N4$) and 
narrow range of MIR colour ($S7-S11$).  The sources classified as  `galaxy' are  mostly distributed in 
the range of  $-1 < N2-N4 < 0$.  The lower right panel  shows the $N2-N4$ colour  as a function of 
redshift.  Figure 2 shows that  MIR colours are efficient tool which can be used to demarcate between 
galaxies and AGNs.   Especially, the  `galaxies' types (which we use  in this work) are gathered at
lower (mostly $z < 0.5$)  redshifts and  in a narrow range of NIR colour ($-1 < N2 - N4 < 0$),   
while having a wider  distribution in the MIR colour domains ($S7 - S11$ and $S11 - L15$) 
suggesting the various kinds of spectral energy distributions (SEDs) in the MIR wavelengths.
Also, there exist intriguing sources such as those classified as `galaxy' but having red NIR colour 
(e.g., $N2 - N4 > 0$),  which we should  investigate further later  when we classify entire sources 
in the NEP-Wide catalogue.

In Figure 3, the number distributions of the entire photometric (dashed) and  spectroscopic  sample (solid) 
are shown in the bottom panels  while the ratio of the spectroscopic sample is shown in the upper panels, 
as a function of observed MIR bands ($S9W$, $S11$, and $L15$) and optical $R$-band (the rightmost
panel) magnitudes. The fraction of spectroscopic sample varies over the MIR magnitudes, and shows broad 
peaks at around 16 (15.5 -- 16.5) magnitudes (AB).  Since the number of sources in optical band is 
much larger than those in the MIR bands, the fraction at $R$-band  is much lower than in the MIR bands.  
In the lower right panel,  a dotted line represents the distribution of R-band sources except for 
the star-like sources defined by the criteria based on the optical mag and NIR colours (Jeon et al. 2010;
Kim et al.  2012; Jeon et al. 2014).  For the  spectroscopic sample, the variation from the darkest to 
brighter shades show  relative amount of sample at the redshift bin  $0.0 < z < 0.1$,  $0.1 < z <0.2$, 
 $0.2 < z < 0.3$, and so on.   Vertical dot-dashed lines indicate the magnitude cuts  used for the 
 spectroscopic target selection.  There are many sources fainter than this criteria  due to the deeper 
 limit of the Hectospec. The incompleteness is more significant at higher redshift bin.  
Our limitation of the redshift range  ($z < 0.3$)   effectively minimize  the incompleteness  
arising from the flux limited sample  by the magnitude cuts.

\subsection{SED Fitting}

Galaxy emission in the MIR is the sum of the Rayleigh-Jeans tail of stellar emission, emission from heated
dust, and a power-law component of AGN, if it exists. Since our interest is to investigate the star-forming
activity and to construct their LFs, we have to know the type of IR sources and classify them into galaxies
or AGNs.    Shim et al. (2013) identified the types of spectroscopic sources  and classified them  into 
star-forming galaxies, starburst-AGN composites, and AGNs based on the line analysis. 
For this work,  we adopted sample classified as `galaxy' having  reliable spectroscopic redshift. 
After excluding AGNs and unreliable spectroscopic redshift, we have 1090 galaxies.  
By limiting the redshift range of galaxies to  $z < 0.3$  to construct the local luminosity
functions, we are left with about $\sim$  600  galaxies.

\begin{figure*}
\includegraphics{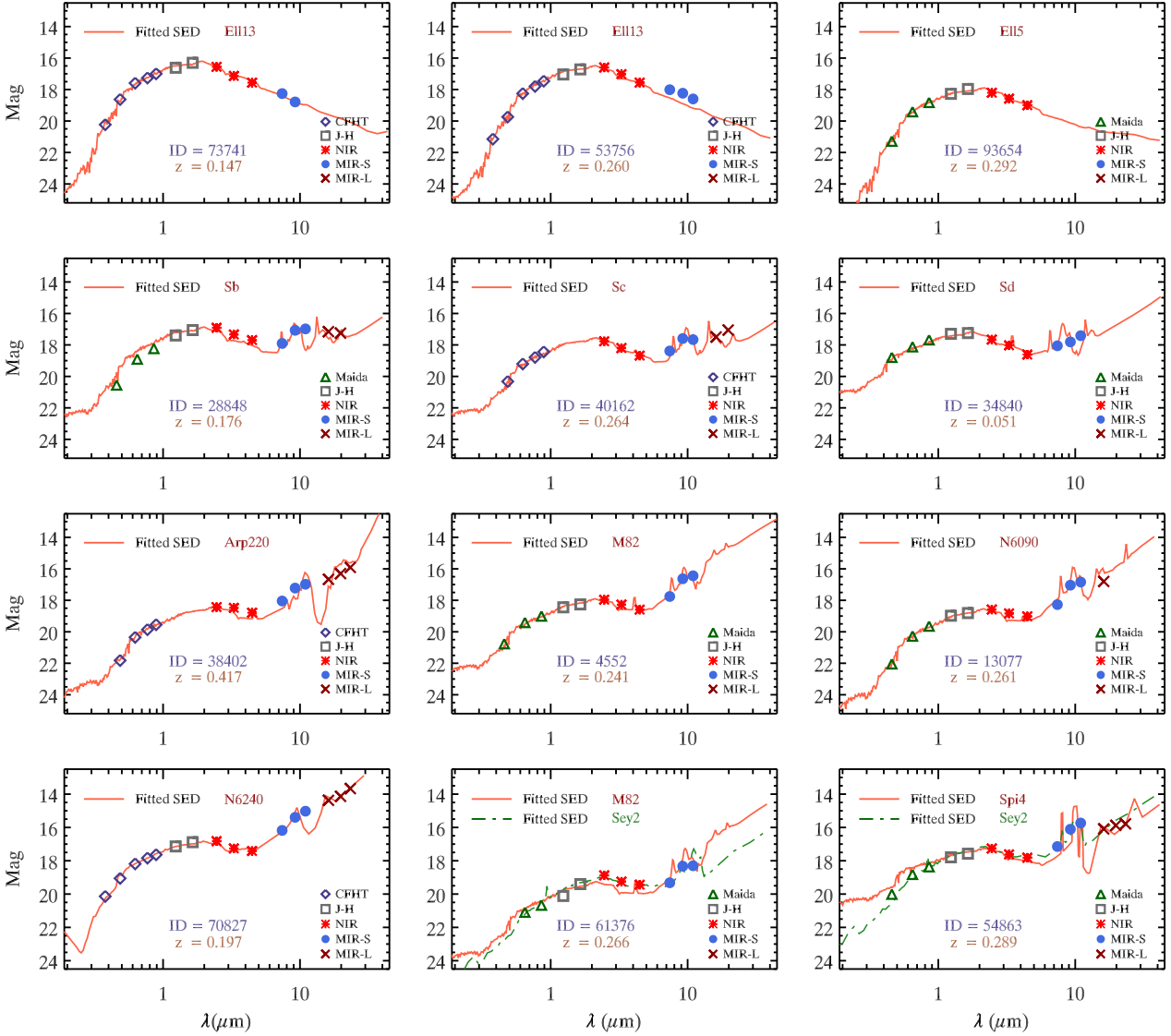}
\caption{ Examples of SED-fitting results for our sample. The horizontal axis is observed  wavelength 
and the vertical axis is observed magnitudes (AB). The small squares indicate available photometric data
points from $u$, $g$, $r$, $i$, $z$ (CFHT: diamonds), $B$, $R$, $I$ (Maidanak: triangles), $J$, 
$H$ (KPNO: squares), $N2$, $N3$, $N4$ (\textit{AKARI} NIR bands: asterisks), and  $S7$, $S9W$,
$S11$ (MIR-S: filled circles), $L15$, $L18W$, $L24$ (MIR-L: x-symples).  Here we show sample galaxies 
fitted to various model templates (e.g., elliptical, spiral, and starburst).  Red lines represent the galaxy 
template from Polletta (2007). In the bottom panel, two galaxies are fitted to both normal galaxies and
Seyfert 2. We have chosen normal galaxy fit in such circumstances because the AGNs  have been
already excluded based on line diagnostics.
\label{fig 4}}
\end{figure*}

 To estimate MIR luminosities for our sample with the spectroscopic-$z$, we carried out SED fit analysis.
We utilized the publicly available code \textit{Le PHARE} (Ilbert et al. 2006). The fitting to SED models 
was done over all the available photometric band data with the spectroscopic redshift (by an option 
ZFIX$=$yes).  We selected an option for the Calzetti extinction law (Calzetti, 2001; Fischera et al.
2003) in the software (EXTINC\_LAW$=$calzetti.dat). Details of the software are given in Ilbert et al. 
(2006).   For each observed SED, \textit{Le PHARE} carried out   single fit  from the $u^{*}$ to the 
mid-IR band,  using various templates defined throughout this wavelength range (Polletta et al. 2007; 
Ilbert et al.  2009).  We searched for a best-fit model among all types  and chose a template giving a 
minimum $\chi^2$.  In Figure 4,  we present some of our local ($z < 0.3$) galaxy sample (e.g., ID$=$ 
73741, 53756, 93654, 28848, 40162, 34840, 38402, 4552, 13077, 70827, and 61376) fitted 
to various templates. The small squares indicate available photometric data points from $u^{*}$, 
$g^{\prime}$, $r^{\prime}$,  $i^{\prime}$, $z^{\prime}$ (CFHT: diamonds), $B$, $R$, $I$ (Maidanak:
triangles), $J$, $H$ (KPNO: squares), $N2$, $N3$ ,$N4$ (\textit{AKARI} NIR bands: asterisks), $S7$,
$S9W$, $S11$ (MIR-S: filled circle), and $L15$, $L18W$, $L24$ (MIR-L: x-symbols). The first row 
shows galaxies that are fitted to  ellipticals (Ell5 and Ell13), the  second row shows  sample  fitted to spiral 
galaxies (Sb, Sc, and Sd types), and the third and bottom rows show those fitted to various starburst 
galaxies, ULIRGs, (e.g., M82, N6090, and N6240 type) and AGN  (e.g., Seyfert type 2).  
 Some sample fits better to AGN template which gave a smaller $\chi^2$  than that of a best-fit galaxy
template, although we used objects  classified as various types of  `galaxy' based on the optical line
analysis.  In this case, both template showed similar acceptable fit over the entire data points  (see the 
last sample presented in Figure 4).  Considering rather complex MIR features and  larger photometric 
uncertainties in the longest part of MIR bands,  slight confusion between galaxies and AGNs for some 
sample seems inevitable.  Because the SEDs from both types are very similar in the MIR, the sources 
fitted better to AGN types do not make significant difference  in our resulting MIR LFs.

\begin{figure*}
\includegraphics[scale=0.8]{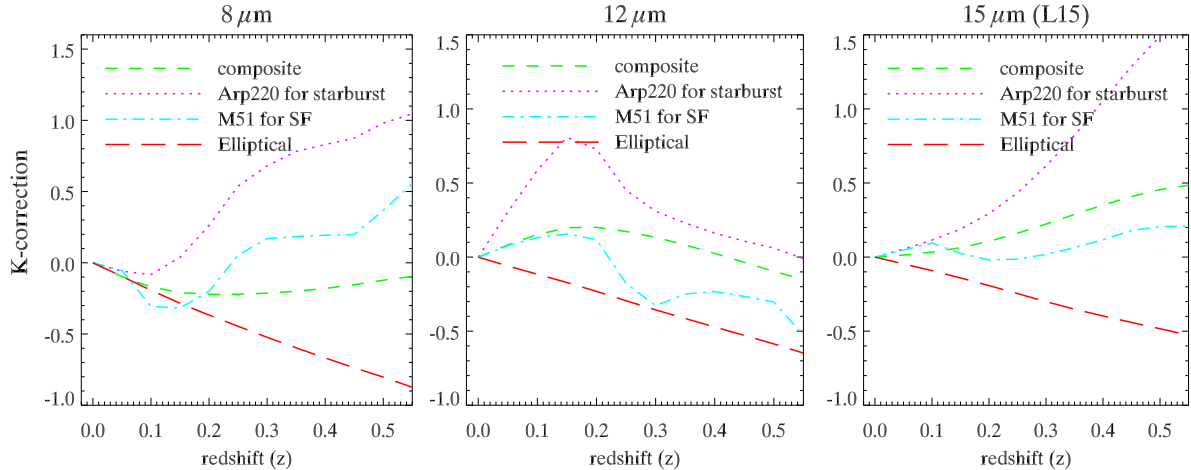}
\caption{K-corrections for   8 $\mu$m, 12 $\mu$m, and 15 $\mu$m 
bands are shown.  The 8$\mu$m  band is strongly affected by the PAH 
emission features around the band widths while the NIR bands take mainly stellar emission. Therefore  
8 $\mu$m bands have the model dependency and shows positive K-correctioin 
for starbursts and star-forming galaxies because those bands include the PAH features around 7 -- 8
$\mu$m in low redshift range.
\label{fig 5}}
\end{figure*}

The K-correction (Oke $\&$ Sandage, 1968), which transforms the measured flux at a given wavelength 
to the unredshifted flux at that band,  is also computed in this procedure. 
We checked the amount of K-corrections for the MIR bands where we derive the LFs,  assuming  simple 
box function centered at 8 and 12 $\mu$m having full-width of 2 $\mu$m  and using L15 band 
response function.  K-correction, depending on the SED of galaxy, its redshift ($z$), and the filter response  
($S(\lambda)$),  can be computed by
\begin{equation}
 K(z) = (1+z)  \frac{\int{ F(\lambda) S(\lambda)} d\lambda}{\int{F({{\lambda}\over{1+z}}) 
 S(\lambda) d\lambda }} .  \\
 \end{equation}
 Figure 5 shows the K-corrections at 8 $\mu$m, 12 $\mu$m, and 15 $\mu$m bands for a few 
 representative SED  models from the Polleta et al. (2007). These figures show substantial differences in 
 the K-correction behavior for different types of galaxies  because  the SEDs in MIR portion vary significantly 
 depending on galaxy types.  For example, the 8 $\mu$m bands show sensitive dependencies, because 
 there are  PAH  features at 6.2 $\mu$m, 7.7 $\mu$m, and 8.6 $\mu$m and some of them can be 
 included or excluded in this band depending on the redshift.   In particular, 8.6 $\mu$m feature is located 
 near the boundary of the 8 $\mu$m band, and thus  can be excluded for the galaxies at low but non-zero  
 redshift.   From z $\sim$ 0.11,  the 7.7 $\mu$m  PAH feature begins to move out of this band. 
 The K-correction is small but not negligible in the MIR bands  for nearby galaxies. 
 In the case of early-type galaxies,  K-correction decreases monotonically with  redshifts 
 since the flux  decreases with wavelength near these  bands.  Our  sample used for this work is 
 located at $z < 0.3$ and  mostly appears to be normal spiral/SF galaxies (strongly dominated by
 late-types), explainable with M51 type (cyan dot-dashed lines in Figure 5), while actively star 
 forming populations such as Arp 220 (or LIRG/ULIRGs) or AGN types appear to take a very
small faction.

\section{Derivation of Luminosity Functions}    

\subsection{$1/V_{max}$ Method}

Luminosity function $\Phi(L)$ is the number density of galaxies  having luminosity in the range $L \sim 
L+\Delta L$  and residing in a specific volume. Therefore, we should take a volume limited sample and 
count all the galaxies in a given luminosity bin and divide it by the surveyed volume. However, it is not as
simple since the sample is usually limited by  flux rather than volume. It means that the survey volume
depends on the brightness or luminosity of the objects in the observation. In order to account for this 
simple fact, $1/V_{max}$ method (as described by Schmidt, 1968) is widely used to obtain the LFs of
flux limited sample.  This method is known to be relatively insensitive to the incompleteness of the 
observations, and there is no parametric dependence or assumed model.   One can calculate  luminosity
($L_{\nu}$) in the rest frame or absolute magnitude ($M_{\nu}$) based on  a redshift and observed 
SED (observed magnitudes $m_{\nu}$ at each band) of each source,  by following formula, 
\begin{equation}
M_{\nu} = m_{\nu} - 5~\log(d_{L}) - 25 - K(z,\lambda), \\
\end{equation}
where $d_{L}(z)$ is the luminosity distance (in units of Mpc) and K is $K$-correction. Using this equation,
we can compute $z_{max}$, which is the maximum redshift at which a source could be observed with the 
detection limits of the survey (at which $m_{\nu} = $ detection limit).   When converting the observed 
flux (or magnitude) in observed wavelength to luminosity of the rest-frame wavelength, 
we need to apply $K$-correction, which is calculated  using the best-fit  template for each 
source (sec. 2.2).  A comoving volume $V_{max} = V(z_{max}) - V(z_{min})$ associated with a 
source is the maximum volume corresponding to the maximum redshift within which the source could be
still detected, where $z_{min}$ is the lower limit of the redshift bin. If $z_{max}$ is larger than the upper
limit of the interested redshift bin (i.e., $z = 0.3$), we take smaller one, i.e.,  $z_{max}$ = min 
($z_{upper}$ of the z-bin, $z_{max}$ of a source).   We estimated  $z_{max}$ and  $V_{max}$  
at each band  where we derive  LFs.  We collected the  sources having 
same luminosity ranges (i.e., finite bin size,  $\Delta L$).  After assigning $V_{max}$ to each source, 
the LF($\Phi$) can be obtained by
\begin{equation}
\Phi(L)  =  \frac{1}{\Delta{L}} ~\sum _{i}   \frac{1}{V_{max,{i}}}~ s_{i}   
\end{equation} 
where $\Phi(L)$ is the number of objects per Mpc$^3$ in the rest frame luminosity range $L \sim (L 
+ \Delta L$), $s_{i}$ is correction factor to compensate the selection bias or incompleteness for $i$th
galaxy, which will be described in detail in the following section 3.2.

\subsection{Incompleteness and Uncertainties}

Since our sample is not  a complete set and has photometric uncertainties, we should take these facts 
into account in deriving the LFs.   First,  the number of galaxies with spectroscopic redshifts are much 
smaller than that of parent photometric sources in the NEP-Wide field, as shown in Figure 3. We compared
the number distribution of the spectroscopic sample with that of entire S11 sources from the NEP-Wide
catalogue as a function of magnitude to obtain the selection function for spectroscopic  sample
because the targets for the spectroscopic observations are selected based mainly on the S11 magnitude. 
In Figure 6, the orange solid line indicates all the S11 sources from the \textit{AKARI} NEP-Wide survey.  
The gray lines show the number distribution of the sources from spectroscopic observation: from top to
the bottom, all the spectroscopic sample,  sample with reliable  redshifts, and those classified as galaxy. 
The solid brown line shows the sources classified as galaxies,  which are used for the estimation
of LFs.   The uncertainty associated with the correction for the number of the spec-z sample can be 
explained by the Poisson statistics in the number of the parent photometric and spectroscopic sample. 
It is normally a few \% level at $S11 > 15.5 $ mag (AB), but reaches around $\sim 10\%$ at 
$S11 < 15 $ mag due to the smaller number of sample. This correction uncertainty appears
to affect  error-bar sizes of LF  by  several \% on average ($\sim 10\%$ at most). 

\begin{figure*}
\includegraphics[scale=0.8]{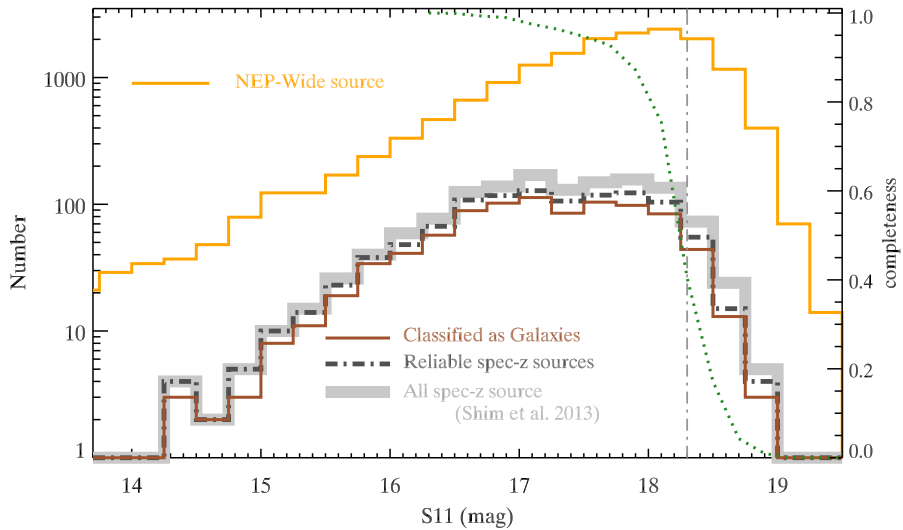}
\caption{The number of sources as a function of observed magnitude at 11$\mu$m band (S11 
band). The orange solid line represents the entire photometric sample of 11$\mu$m sources from the 
\textit{AKARI} NEP-Wide photometric survey. The gray lines show the number distribution of the 
spectroscopic sample: from top to the bottom, all the spectroscopic-$z$ sources (thick white-gray)  
and spectroscopic sample with reliable redshifts (dot-dashed dark gray). The solid brown line shows the distribution of sample classified as `galaxy' among 
the sample with reliable redshifts.  A vertical dot-dashed line indicates the primary selection criteria for the
 spectroscopic observation ($\sim$ 50$\%$ completeness limit). Dotted line is the photometric
incompleteness based on artificial star simulations by Kim et al. (2012).
\label{fig 6}}
\end{figure*}

Second, the parent photometric data becomes partially incomplete near the detection limit.  At the MIR-S
 (S7 -- S11) bands, the detection limit ranges 19.5 -- 19 mag  and  the 50\% completeness levels  
are  $\sim$ 0.6 mag  brighter than the detection limits.  The magnitude difference between the 90 $\%$
and 10 $\%$ completeness is about 0.7 -- 0.8 mag, indicating that the completeness begins to drop 
quite rapidly from around  90 $\%$ level (18.3 -- 17.9 mag).  We should also be careful at the bright 
end  because the low sampling rate could derive unnatural selection function.   In Figure 6, the green dotted 
line with the right-side vertical axis  indicates the completeness level at S11 band  obtained by simulation  
of  artificial source injection and re-extraction tests  as described by Kim et al. (2012).   A vertical 
dot-dashed gray line  roughly corresponds to the primary selection criteria  for the  spectroscopic 
observation (50$\%$ completeness limit of the NEP-Wide survey).  We applied the correction for 
this photometric incompleteness  based on the completeness estimation of  the  $S11$ data that was
obtained by Kim et al. (2012) in our derivation of the LFs.  There is also redshift bias as shown in the 
right-hand panel of Figure 1.  We compared the number density of spectroscopic sources in the comoving
volume corresponding to each redshift bin in order to estimate the redshift incompleteness.  Assuming
constant number density, we estimated the correction for incompleteness as  a function of $z$.  
Since our derivation of LFs is limited to $z < 0.3$  where the source distribution is roughly consistent 
with the constant spatial density,  we regard the redshift bias is not significant.

Rest frame luminosity is directly derived from SED-fit results, therefore, error estimation for luminosity
function (LF) depends on the SED-fitting errors arising  from   redshift, $k$-correction error, and 
photometric uncertainties. Since we used the redshifts measured by spectroscopic observation,  we can 
ignore uncertainties from the redshifts (which is very small). Thus, our LF errors can be determined based
mostly on the photometric error.   In order to estimate the errors of the LFs, we carried out Monte Carlo
simulation  using random number generator.   We generated more than 10,000  simulated catalogues by
putting random errors into the measured flux of each source. In the simulated catalogues, those errors are
realized by assuming a normal Gaussian distribution centered at the observed flux.  Standard deviation 
of this gaussian distribution for simulated errors  is determined based on photometry measurement error.   
This means we generated $\sim$ 10,000 input catalogues for Le  PHARE runs doing SED-fit and 
obtained result sets, which finally results in  slightly different 10,000 LFs.  
We determined final  luminosity function and its uncertainties  based on weighted
average and the dispersion of the simulated results.

\section{Results and Discussion}

\subsection{8 $\mu$m Luminosity Function}

It is known that  8 $\mu$m luminosity of  galaxy is well correlated  with the total IR luminosity (Babbedge 
et al. 2006; Caputi et al. 2007; Bavouzet et al. 2008; Goto et al. 2010; Galametz et al. 2013) 
because the rest-frame 8$\mu$m fluxes are dominated by prominent PAH features which are sensitive
to star formation activity.  In this section, we present 8 $\mu$m luminosity function of local galaxies in 
the NEP-Wide field.  To estimate 8$\mu$m  fluxes at observed frame, we used  a simple boxy filter
centered at 8$\mu$m with a 2$\mu$m bandwidth across the filter (constant transmission over the 
bandwidth)  and  convolved  the filter response  with best-fit template for each galaxy. 
 The observed flux and the rest-frame
luminosity are related by  following equation
\begin{equation}
 L_{\nu}(\nu_{rest} ) = \frac{4{\pi}{D_{L}}^{2}} {(1 + z)}  F_{\nu}(\nu_{obs}),
\end{equation}
where $D_{L}$ is the luminosity distance for a given redshift based on the cosmological parameters
assumed in this work, $F_{\nu}$ the flux density (erg cm$^{-2}$ Hz$^{-1}$), and $\nu_{obs}$ and 
$\nu_{rest}$ are the observed and rest-frame frequencies, respectively, related by $\nu_{rest} =
 (1+z) \nu_{obs}$.

\begin{figure*}
\includegraphics[scale=0.9]{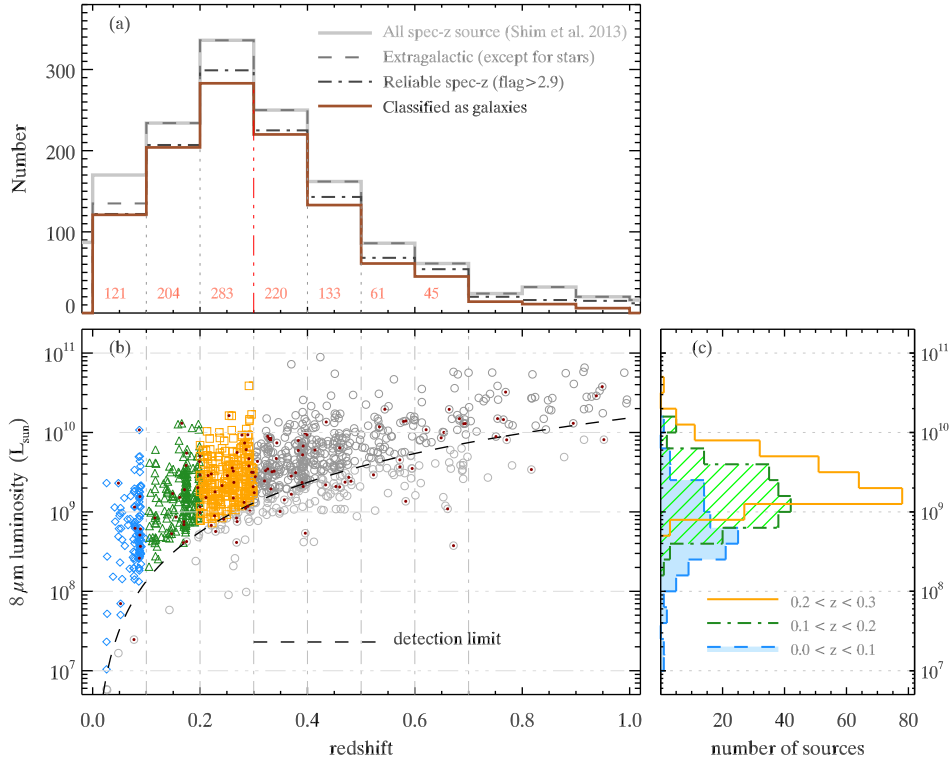}
\caption{ Top panel (a) shows the fraction of the sources which is classified as galaxies among all the
spectroscopic sample as a function of redshift. Solid gray line indicates all the spectroscopic sources, 
dotted-dashed lines represents the sources with the reliable redshifts, and the black line shows the number 
of sources classified as galaxies. At the bottom, left panel (b) shows the luminosity distribution of galaxies 
at rest-frame 8 $\mu$m as a function of redshift. Dark red points represent sample classified as AGN.
A broken curve indicates the flux limit of observation 
at 8 $\mu$m according to redshift. Bottom right panel (c) shows the number of galaxies with 8 $\mu$m 
luminosity in each  redshift bin from 0.0 to 0.3. Sky-blue colour represents the sample in the redshift 
$z = 0.0 - 0.1$, green colour indicates the sample in the range of  $z = 0.1 - 0.2$, and orange colour 
shows the sample in  $z = 0.2 - 0.3$. 
\label{fig 7}}
\end{figure*}

\begin{figure}
\includegraphics[scale=0.6]{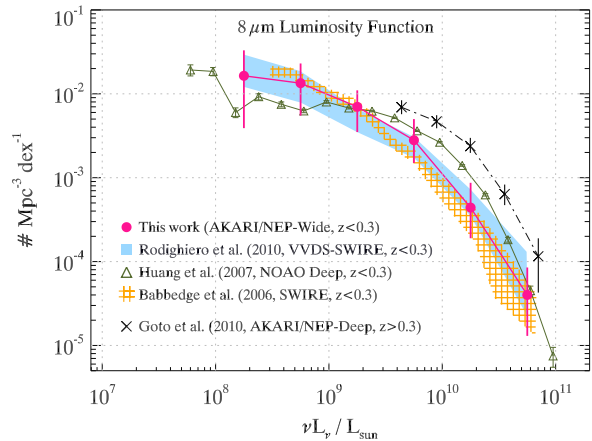}
\caption{  8 $\mu$m luminosity functions (LFs) for local SF galaxies.  The filled circle (magenta colour)
represents the LF of local ($z < 0.3$) galaxy sample in the NEP-Wide survey field (this work).
Shaded area indicates the LF of Rodighiero et al. (2010) using VVDS-SWIRE data. Meshed region shows 
the results from Babbedge et al. (2006) based on the SWIRE data. Triangles are from Huang et al. 
(2007, NOAO deep), and  `x' symbols from the G10.
\label{fig 8}}
\end{figure}

In Figure 7, we  show the distribution of the sources used to derive 8 $\mu$m luminosity function. 
The top panel (Figure 7a) shows the number of sources versus redshift. The width of the redshift bin  
is 0.1, as indicated by vertical gray broken lines.  From all the spectroscopic sample to those classified 
as galaxies, the amount of sample  is shown  by solid gray line (all spec-z sample), broken lines (except
for stars and reliable spec-z), and the brown solid line (galaxies).  We give the number of galaxies on  
histogram bar chart for each redshift bin. The bottom left panel (Figure 7b) shows the 8 $\mu$m 
luminosity distribution of sample as a function of  redshift. The squares represent all the spectroscopic 
sample,  but  for $ 0 < z < 0.3$,  the sample in each  redshift bin over the detection limit  is over-plotted with diamonds, triangles, and squares, respectively.  The  dark red dots represent those classified as 
AGNs, and a dashed curve indicates the luminosity corresponding to the 8 $\mu$m flux limit as a 
function of redshifts. 
{The estimation of the flux limit depends on SED model,
but in the range  $z < 0.3$, different models do not make significant changes ($\sim$ a few \%).  
We estimated the 8 $\mu$m flux limit based on a simple flat SED.}   
For the NEP-Wide survey, the detection limits  in the mid-IR  bands of IRC increases almost linearly  
as a function of a filter's central wavelength.  Therefore we derived  detection limit at 8 $\mu$m through the interpolation.  Our sample is brighter than the detection limit of S11 band, but it is not always in 
8 $\mu$m band.   Since 8 $\mu$m flux and luminosities are estimated from the SED-fit results,  
some sample can have 8 $\mu$m luminosities lower than the detection limit curve.  In the bottom right 
panel (Figure 7c), the number distribution of these sources are shown as a function  of 8$\mu$m 
luminosity for each redshift bin using the same colour as used in the left panel.

 We constructed   the 8 $\mu$m luminosity function for star-forming galaxies (at $0 < z < 0.3$) using 
the sample presented in the Figure 7c.  In the redshift bin of $0 < z < 0.1$, a vertical clustering (Figure 7b) 
at around z $\sim$ 0.09 is due to the presence of a supercluster at z $=$ 0.087  in the NEP-Wide 
survey data. We did not include it ($\sim$ 90 sample at $0.07 < z < 0.1$, see Ko et al. 2012)  
for the estimation of luminosity function because its local over-density could affect the luminosity function. 
We present the 8$\mu$m luminosity function for
local SF galaxies (SFG) in the NEP-Wide field in Figure 8.   While a large fraction of the sample occupy 
$10^{9}$ -- $10^{10}$ luminosity range (Figure 7c),  there are too small number of sources
fainter than $10^{8} L_{\odot}$.  At around $z \sim 0.1$,  the 8 $\mu$m detection limits according 
to the redshift exceeds   $10^{8} L_{\odot}$ indicating  many local sample fainter than $10^{8} 
L_{\odot}$  could not be detected in this survey. Therefore, it is hard to produce a statistically meaningful
$\Phi(L)$ for this  luminosity range. 
We compared our LF with previous results from various literatures.   In Fig 8,  shaded area indicates 
the LF from  Rodighiero et al. (2010), which is based on the data from \textit{Spitzer} surveys 
on the VIMOS VLT Deep  Survey (VVDS-SWIRE) and GOODS fields.   Triangles  show the LF of local 
($z < 0.3$) galaxies in the NOAO deep field  in Bo\"otes field (Huang et al., 2007).   Meshed area 
indicates the LF of local ($z < 0.25$) galaxies in the SWIRE field from Babbege et  al. (2006). 
These different works are summarized in Table 1. 
Rodighiero et al. (2007) used a library of template SEDs based on Polletta et al. (2007) and 
Franceschini et al. (2005).  Babbedge et al. (2006)  used the SED models from  Rowan-Robinson 
et al. (2005)  and photometric-redshifts.  Their SFG/AGN separation is based on  SED-fit results.  
Huang et al. (2007) used models from Lu \& Hur (2000) with spectroscopic redshifts.   
A part of works from G10 presented here shows a LF for a higher redshift range ($0.38 < z < 
0.58$), which was based on  the  NEP-Deep survey and  photometric redshifts ($\sim$ 500 
sample).  We fit our 8 $\mu$m LF  using 
 a double -power law (Marshall 1987; Babbedge  et al. 2006;  Goto et al. 2010) as
 \begin{equation}
   \Phi(L) {{dL} / {L^{*}}}  = {\Phi}^{*} \left[ \left ({{L} \over {L^{*}}} \right)^{1-\alpha} +
   \left ({{L} \over {L^{*}}} \right)^{1-\beta}
   \right] { {dL} \over {L^{*}}}  .
\end{equation}
The best fit  parameters for the normalization factor ($\Phi^{*}$), and the characteristic luminosity 
($L^{*}$)  are  $2.2\times 10^{-3}$ (Mpc$^{-3}$dex$^{-1})$ and $4.95 \times 10^{9} 
(L_{\odot})$, respectively.  We obtained the faint-end slope $\alpha = 1.53$, and the bright-end slope 
$\beta = 2.85$, which are comparable to those   of the adjacent  redshift bin ($0.38 < z < 0.58$) from
G10.  Our  local   LF  spans wide luminosity ranges and  is clearly shifted toward the lower density at a 
given luminosity in comparison with the LF from G10, indicating luminosity evolution.  
Our LF shows a good agreement with that of Rodighiero et al. (2010) but, in general, is consistent with
other works.   The difference between LFs also exist (e.g., some deviation of Huang's).  
At the faintest luminosity ranges,    rather larger photometric errors (with smaller number statistics if any)  
can count for the fluctuation. At higher luminosity ranges, it is more difficult to track down the origin of the 
differences in detail.  But the differences seem to originate mostly from the different incompleteness in each 
sample and corrections for them.   We regard that different number of sources contributing to each luminosity
 bin and AGN fraction to be excluded are also entangled and seem to affect the disagreement, even though
  it's hard to compare how the individual works deal with correction factors.

\subsection{Luminosity Functions for 12 $\mu$m  and 15 $\mu$m }

\begin{figure}
\includegraphics[scale=0.6]{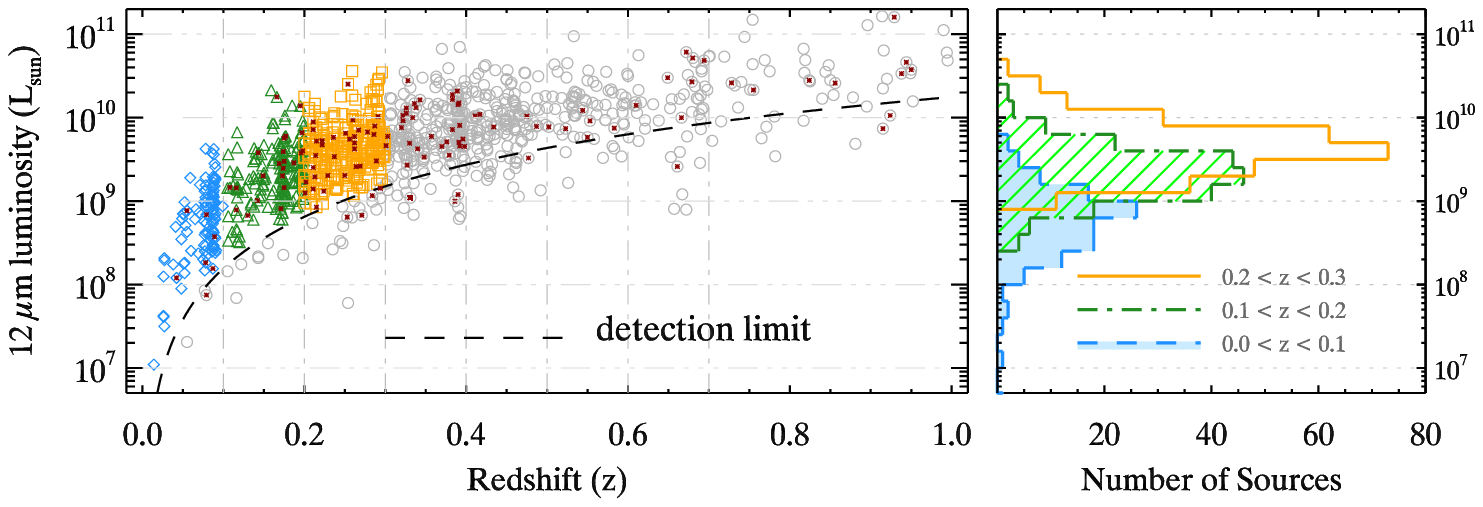}
\includegraphics[scale=0.65]{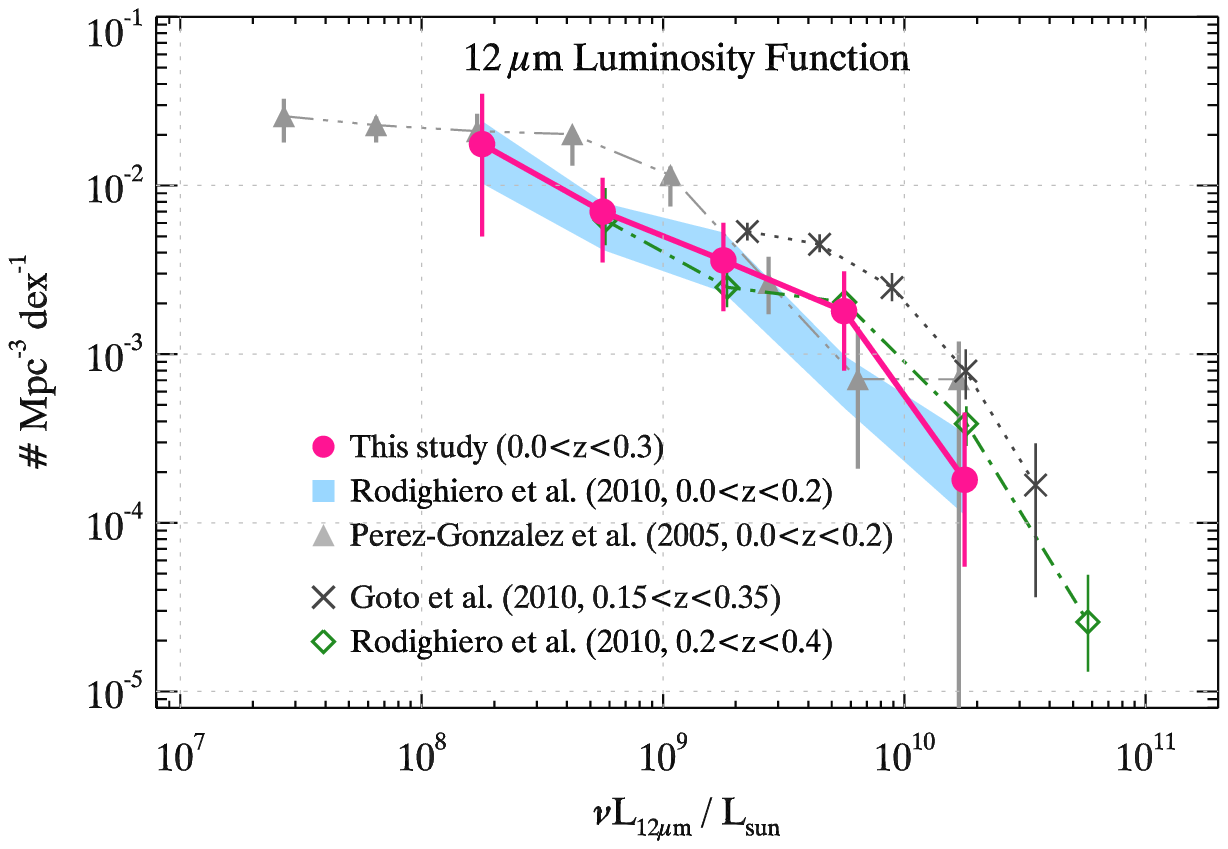}
\caption{ Top panels show 12 $\mu$m luminosity distribution of local  galaxies in the NEP-Wide 
field as a function of redshift (left) and number of sources in each luminosity bin (right). Bottom panel 
shows the 12 $\mu$m  luminosity function derived from the upper panels.  In the bottom panel,  
filled circles represent this work for local ($z < 0.3$) galaxies in the NEP-Wide survey field.
 Triangle represent the local LF derived  by P{\'{e}}rez-Gonz{\'a}lez et al. (2005).
Shaded area and diamonds indicate the LF for $0.0 < z < 0.2$ and  $0.2 < z < 0.4$  from
Rodighiero et al. (2010) using VVDS-SWIRE data.  `x' symbols indicate LF for galaxies at
$0.15 < z < 0.35$ from the G10.
\label{fig 9}}
\end{figure}

We construct 12 $\mu$m  luminosity function (LF) of local galaxies similar way.  12$\mu$m 
luminosity has also been studied well based on the IRAS and ISO observations (Spinoglio et al. 1989;
Rush et al. 1993). It is also known  to have good correlation  with total IR luminosity, and recognized 
as one of the indicator of galaxy SF activity. (Spinoglio et al. 1995; P{\'{e}}rez-Gonz{\'a}lez et al. 
2005). The continuous wavelength coverage of our photometric data (from the near-IR) all the way
through  24 $\mu$m band is very useful/helpful to obtain accurate 12 $\mu$m  luminosity based on
 good SED fit with spectroscopic redshift.  The detection limit at 12$\mu$m band corresponds to about 
18.8 mag and  50\% completeness level is around 18.0 mag (AB).    Due to the slightly different limit
and luminosity distribution  from those of the 8 $\mu$m,  the number of sample selected in each 12
$\mu$m luminosity bin differs a few $\%$ from  that  of  8 $\mu$m  luminosity function (LF),  as 
shown in the upper panels of Figure 9.

In the lower panel of Figure 9, we present  our 12 $\mu$m luminosity functions (for $z < 0.3$)  along 
with other results.   The filled circles (magenta) represent the LF of local ($z < 0.3$) galaxies in the  
NEP-Wide field (this work).   The shaded area (in light blue) and diamond (green) symbols represent the
 LFs  from Rodighiero et al. (2010) based on the VVDS-SWIRE and GOODS data,  for the redshift
$0.0 < z < 0.2$ and  $0.2 < z < 0.4$,  respectively.  The filled triangle indicate the LF determined by 
P{\'{e}}rez-Gonz{\'a}lez et al. (2005) for $0.0 < z < 0.2$. We also compared with the 12 $\mu$m
 LF derived by G10 for galaxies at  $0.15 < z < 0.35$ from the NEP-Deep data.  
The redshift ranges from these various works are partly overlapped with each other, but the filled circles,
triangles and shade area represent the LFs for the local universe,  while the diamond and cross symbols correspond to the LFs of higher redshift ranges.  Our result (filled circles) for local galaxies agrees well with the 
Rodighiero's local LF,  except for a deviated point at the luminosity bin $ 9.5 < \log  
{\nu}L_{\nu(12{\mu}m)} < 10$,  seemingly due to the different  redshift ranges.   For the higher-z 
LFs,  the redshift  range $0.2 < z < 0.4$ (diamonds) of Rodighiero et al. (2010) is slightly higher than
that from G10 ($0.15 < z < 0.35$),  but their LF is systematically  lower than that of Goto et al. 
(2010).  
Both works are based on the photometric redshifts, thereby seem to contain the inevitable uncertainties. 
Besides, due to the samplings by different redshift bins,  probably leading the different correction schemes
for each incompleteness/bias might have caused the deviation from each other.  However, the 
measurements of LFs for both fields (the NEP and the VVDS-SWIRE field) consistently suggest the
evolution  of 12$\mu$m LF.

\begin{figure}
\includegraphics[scale=0.6]{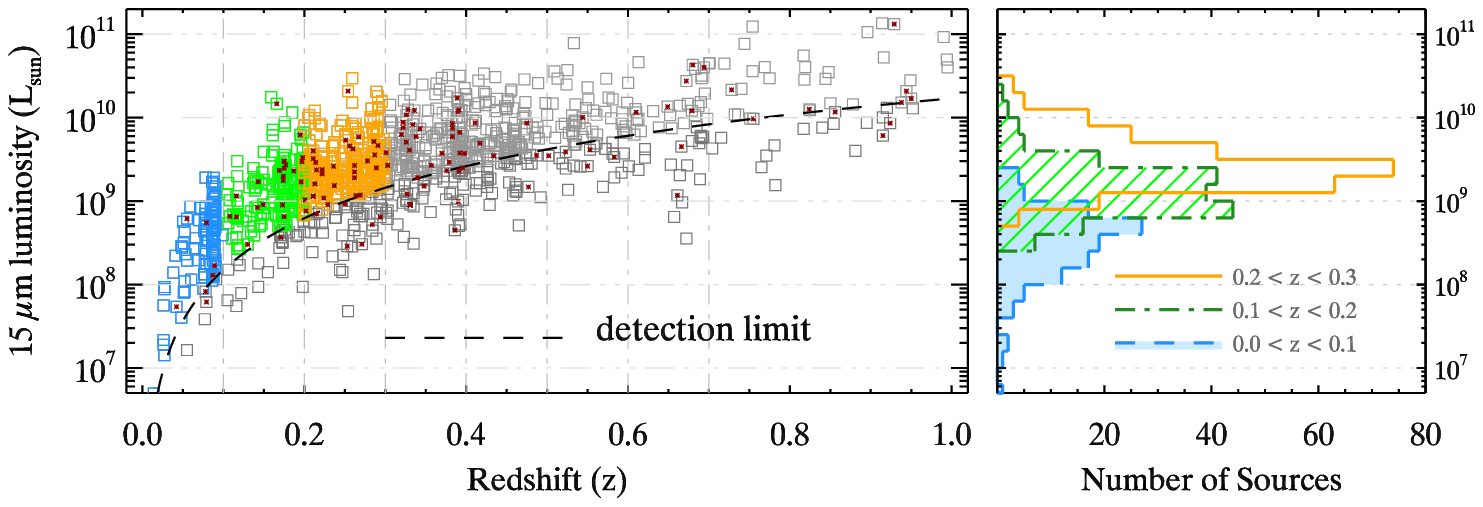}
\includegraphics[scale=0.65]{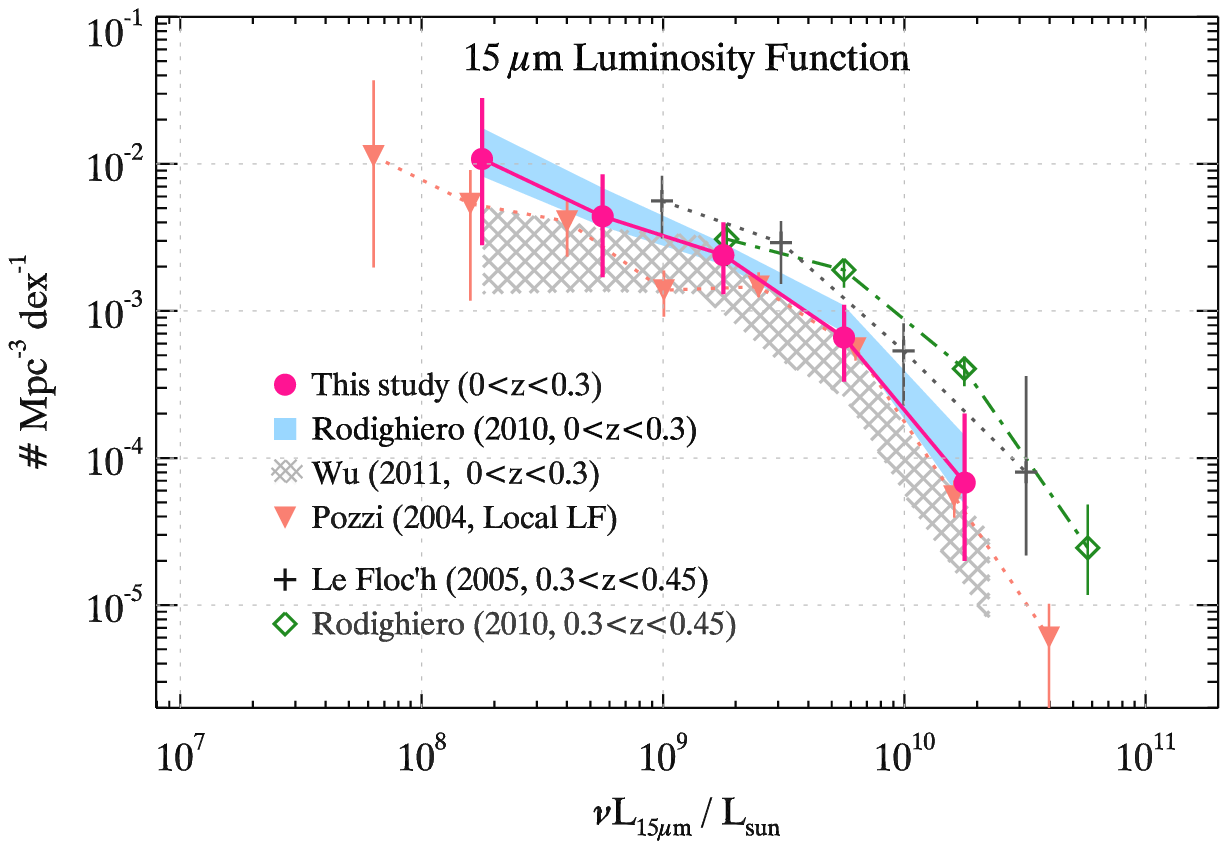}
\caption{Top panels show 15 $\mu$m luminosity distribution of local SF galaxies in the NEP-Wide 
field as a function of redshift and number of sources in each luminosity bin. Bottom panel shows the 
15 $\mu$m  luminosity function derived from the upper panels.  In the bottom panel, the circles 
(magenta colour) represent  the LF of our ($z < 0.3$) sample in the NEP field. 
Triangles indicate the local LF from Pozzi et al. (2004),  and shaded area indicates the local LF
from Rodighiero et al. (2010). Meshed region  represents the results given by Wu et al. (2011).
Also local LF from Le Floc'h (2005) was  presented by crosses.
\label{fig 9}}
\end{figure}

We  also  derive 15$\mu$m luminosity function of local ($0.0 < z < 0.3$) galaxies in the NEP-Wide 
field and compare with various other results. We used the same galaxy sample as used for the 8 
$\mu$m  and 12 $\mu$m LF,  but slight different number of sample contributes to each luminosity bin
(see Figure 10).  While  the observed magnitudes (or fluxes) at 8 $\mu$m  and 12 $\mu$m bands  
derived from the assumed simple boxy filters with fitted SED templates, the observed  15 $\mu$m fluxes
used in this section were based on the measurement by L15 filter of IRC.  The 15$\mu$m LF was 
needed for interpretation of 15$\mu$m source properties (Xu, 2000; Gruppioni et al., 2005) since 
MIR observations by  ISOCAM,   its significance as an indicator to estimate the star formation activity 
has  also been recognized.  Since  Xu et al. (2000) derived local ($z < 0.2$) luminosity function based
on ISOCAM 15$\mu$m observation on the IRAS 60$\mu$m galaxies ($\sim$100 sample) in the NEP
 region, various determination of 15$\mu$m LF has been carried  out.   We present the 15
$\mu$m  luminosity distribution of our sample and the resulting LF  in Figure 10  in comparison with 
several  previous works. Here, the filled circles (magenta) represent our results ($z < 0.3$).   
Pozzi et al. (2004) determined the local ($0 < z < 0.2$ and $0.2 < z < 0.4$) 15$\mu$m luminosity 
function using  about 
150 sample from the European Large Area ISO Survey (ELAIS) southern  (S1: 2$^{\circ}$$\times$ 
2$^{\circ}$, S2: $21^{\prime}\times 21^{\prime}$ ) field, based on the  SED analysis mainly on 
M51 and M81 types as reference templates. They did not include active galaxies like Arp 220 ($L_{IR}
 > 10^{11} L_{\odot}$) considering that those types are rare and tend to appear at higher-z (e.g., 
 $z > 0.8$).  This aspect seems consistent with our local spectroscopic sample  but we used many 
 types of  normal galaxy templates  in the SED fit analysis.  
Wu et al. (2011) determined the LF for the redshift range of $z < 0.3$ based on the galaxies ($\sim$
230)  from the 5 Milijansky Unbiased Extragalactic Survey (5MUSES), with the redshifts measured by 
Infrared Spectrograph  (IRS). Their LF is also presented  by meshed area in Figure 10. 
There is no 15$\mu$m LF measurement  for the higher redshift from the NEP-Deep data.  But the 
LFs  from Le Floc'h  et al. (2005) and  Rodighiero  et al. (2010)  indicated by cross and diamond 
 represent higher-z ($0.3 < z < 0.45$) LFs .  Here, Le Floc'h et al (2005) used \textit{Spitzer} MIPS 
 24$\mu$m selection at  CDF-S field (0.6 deg$^2$),  carrying out analysis about 2600  sample 
 based on both spectroscopic  and photometric data from literatures and COMBO-17.
 These various works are also summarized in Table 1.

 \begin{table*}
\begin{center}
\caption{Summary table showing LFs from  a number of different works.}
\begin{tabular}{llccrllclll}
\hline\hline
    LF works  & {Survey  Field} & Size of area & Band & 
 \multicolumn{1}{c}{N$_{source}$} & &\multicolumn{2}{c}{Redshift   info} &   \\
    (authors)  & &[deg$^{2}$] &  [$\mu$m]    &     & & (range) & (spec or phot)  \\
\hline 
 This work                             & NEP-Wide      & $\sim5.4 $ & 8, 12, 15  &  $\sim ~600   $ &  & $ z < 0.3$   &  spec-z &  \\
 Babbedge et al. (2006)        & ELAIS N1      & $\sim6.5 $ & 8               &  $\sim 5,000$ &  & $ z < 0.25$ &  phot-z &  \\
 Huang et al. (2007)             & NOAO Deep  & $\sim6.8 $ & 8               &  $\sim 2,600  $ &  & $ z < 0.3 $  & spec-z  &  \\
 Rodighiero et al. (2010)        & VVDS-SWIRE& $\sim0.9 $ & 8, 12, 15  &   $\sim 440 $ &  & $ z < 0.3 $  &  phot-z  &  \\ 
 P{\'{e}}rez-Gonz{\'a}lez et al. (2005)& CDFS/HDFN  & $\sim0.5 $ & 12   &  $\sim   500$ &  & $ z < 0.2 $  &  phot-z  &  \\
  Pozzi  et al. (2005)           & ELAIS South  & $\sim0.5 $ & 15                &  $\sim   150$ &  & $ z < 0.2 $  &  spec-z  &  \\
 Wu et al. (2011)                  & 5 MUSES      & $\sim40. $  & 15            &  $\sim   230$ &  & $ z < 0.3 $  &  spec-z  &   \\
\hline
 Goto et al. (2010)        &  NEP-Deep       & $\sim 0.6 $ & 8, 12    &$\sim  500 $ & &  $0.2<z<0.6$   &  phot-z   & \\
 Rodighiero et al. (2010) & VVDS-SWIRE   & $\sim 0.9 $ & 12, 15 & $\sim  300 $  & & $0.3<z <0.45$ &  phot-z &  \\
 Le Floc'h et al. (2005 ) &    CDFS           & $\sim 0.6  $ &    15     & $\sim 2,600$  & & $0.3<z<0.45$  &  phot-z  &  \\
\hline\hline
\end{tabular}
\end{center}
\label{tab:Comparison_LF_works}
\end{table*}

\section{SUMMARY   AND CONCLUSION}

We presented mid-infrared luminosity functions for the galaxies in the local universe.  We used the 
sample from the NEP-Wide survey data observed by AKARI space telescope, which covered the near- 
to mid-IR (2 $\sim$ 25 $\mu$m) wavelengths continuously.  To obtain more accurate luminosity 
functions, we took advantage of the spectroscopic redshift information obtained by optical  follow-up 
surveys carried out with the MMT/Hectospec and WIYN/Hydra as well  as various ancillary data sets
covering from optical u$^*$ to NIR $H$ band.  Therefore the SED-fit analysis to determine the rest 
frame luminosities at MIR bands does not have serious source of uncertainties.  For better statistics 
and completeness, we limited this study to the lower redshifts ($z < 0.3$) range. 
Our sample appears to be composed of various types of local SF galaxies including a small fraction 
of ellipticals and normal galaxies with some of these having similar SEDs to those of active galaxies.
For the purpose of this work to derive the MIR luminosity functions at the 8, 12, and 
15 $\mu$m bands, we employed the $1/V_{max}$ method, which is widely accepted 
because  of its merits as described in sec. 3.1.  
 
The LFs appear to be consistent with other previous studies, in general, within error bars although the 
differences among various determinations seem to originate from  different method for source 
classification/selection and different photometric data and SED models, different redshift measurements 
and redshift ranges of sample, different correction methods for incompleteness or selection effect,  
different fields/areal coverage, and so on.  Errors also arises from  those components.  But, compared 
to the other works,  the advantage of this work is the redshift information determined by spectroscopic 
observation and source classification based on line analyses as well as extensive wavelength coverage 
from optical u$^{*}$ to MIR  25 $\mu$m  including the  \textit{AKARI}'s continuous filter coverage
 in the MIR wavelengths, which allow us to obtain good SED-fit without large uncertainties such as 
 those from photometric  redshifts or type decision by colours or SED fit.   Also, since our sample is 
 distributed  over the large area of 5.4 $deg^{2}$, our results are not susceptible to the statistical 
 uncertainty arising from the observation on a small part of the sky.  While AKARI data do not drastically
change earlier results, we can give more accurate luminosity  function for the luminosity ranges 
10$^{8}$ -- 10$^{10}$  L$_{\odot}$.   Although it is not easy to say about evolution  based on 
this work alone,  valuable comparisons  are possible with the higher-z LFs from the NEP-Deep survey  
which  is complementary to  NEP-Wide or various LFs of similar redshifts.  The purpose of this work 
was achieved   but  exact measurements for more fainter luminosity bins and faint end slope seem to 
be pending.      We expect to  extend this study to the FIR wavelengths range and to higher redshifts
using much larger number of sample  than those used in this work,
if we can measure photometric redshifts for the sources in the NEP-Wide catalogue.

\section*{Acknowledgments}
We would like to thank the referee for the careful reading and constructive comments for this paper.   
This work is based on observations with \textit{AKARI}, a JAXA project with the participation
of ESA, universities and companies in Japan, Korea, the UK, and so on.  
This work was supported by the grant 2012R1A4A1028713 from National Research 
Foundation of Korea (NRFK). 
MI acknowledges the support from the National Research Foundation of Korea grant, 
No. 2008-0060544. SJK thanks Dani Chao and the Institute of Astronomy, National Tsing Hua 
University, for their hospitality during his visit.

\end{document}